\documentclass[oneside,12pt]{article}

    \title{Time and Observables in Unimodular Gravity }

    \author{ Hossein Farajollahi \footnote{School of Mathematics and
    Statistics,
    University of Sydney,
    NSW 2006, Australia,Tel:(612)93515726, Fax:(612)93515435,
    Email: hosseinf@maths.usyd.edu.au  }}

     \date{}

\begin{document}
    \maketitle

    \begin{abstract}

A cosmological time variable is emerged from the hamiltonian
formulation of unimodular theory of gravity to measure the
evolution of dynamical observables in the theory. A set of
'constants of motion' has been identified for the theory on the
null hypersurfaces that its evolution is with respect to the
volume clock introduced by the cosmological time variable.

$\bf{Keywords}$: Relativity, Unimodular, gravity, Time,
Observable,

\end{abstract}

    \newpage

    \def\be{\begin{equation}}
    \def\ee{\end{equation}}
    \def\bea{\begin{eqnarray}}
    \def\eea{\end{eqnarray}}
    \def\M{{{\cal M}}}
    \def\bdy{{\partial\cal M}}
    \def\w{\widehat}
    \def\n{\widetilde}
    \def\real{{\bf R}}

    \section{Introduction}

Research in quantum gravity may be regarded as an attempt to
construct a theoretical scheme in which ideas from General
Relativity and quantum theory are reconciled. However, after many
decades of intense work we are still far from having a complete
quantum theory of gravity. Any theoretical scheme of gravity must
address a variety of conceptual including the problem of time and
identification of dynamical observables. There are many program
that attempt to address the above mentioned problems including
canonical quantum gravity.

It is well know that some of the issues such as time and
observables in quantum gravity have their roots in classical
general relativity; in such cases it seems more reasonable to
identify and perhaps address the problem first in this context.The
classical theory of gravity is invariant under the group of  Diff
(${\cal M}$) of diffeomorphisms of the space-time manifold ${\cal
M}$ . This goes against the simple Newtonian picture of the a
fixed and absolute time parameter. The classical theory, while
itself free from problems relating to the definition and
interpretation of time, contains indications of problems in the
quantum theory, where the absence of a time parameter is hard to
reconcile with our everyday experience. In fact, one can see that
in the hamiltonian formulation of classical general relativity,
time is suppressed from the theory.There are many proposals for
dealing with this question which generally involve a
re-interpretation of the usual notion of time ( see \cite{Isham}
for an overview of these proposals).

 Unimodular gravity as an alternative theory of gravity was
 originally considered by Einstein \cite{Einstein} cast into canonical form
 by Unruh \cite{Unruh1989}and others \cite{Henneaux} for the purpose of constructing
 an explicit time variable for the theory. It contains a pair of canonically
 conjugate fields that are not present in the canonical formulation of conventional
 General Relativity. One of the new fields specifies the value of the cosmological
 constant, while the conjugate field carries the information about the space-time
 volume bounded by the initial and final space-like hypersurfaces.
 The four-volume variable may be regarded as a cosmological time. In fact, in formulating the Einstein
 theory of relativity, one chooses to limit the geometries  by specifying a fixed value
 for the total four-volume.  This produces unimodular theory whose classical
 limit is equivalent to the Einstein theory except that the cosmological
 constant becomes a constant of integration, rather than a
 dynamically unchangeable parameter in the Lagrangian. Limiting the geometries
in this way may solve the timeless character of the quantum
gravity.

 Identification of dynamical observable for the theory is another fundamental issue
that has its roots in classical formulation of general relativity
and directly related to the issue of time. The problem of evolving
of a dynamical system from initial data is known as the Cauchy
problem or initial value problem \cite{Inverno} and in General
Relativity is naturally addressed using the 3+1 ADM
representation. In the ADM approach, the spatial hypersurface
$\Sigma$ is assumed to be equipped with a space-like 3-metric
$h_{ij}$ induced from space-time metric $g_{\mu \nu}$. Einstein's
equations are of course covariant and do not single out a
preferred time with which to parametrise the evolution.
Nevertheless, we can specify initial data on a chosen spatial
hypersurface $\Sigma$ , and if $\Sigma$ is Cauchy, we can evolve
uniquely from it to a hypersurface in the future or past. The
issue of specification of initial or final data on Cauchy
hypersurfaces has been discussed in many papers; for example, see
\cite{Hawking}.

 An alternative approach to Cauchy problem is known as characteristic
 initial value problem in which one may fix the initial data on null hypersurfaces rather than
spatial hypersurfaces. There are reasons to motivate us using null
boundaries in formulating general relativity. First, the procedure
of determining 'initial conditions' on space-like hypersurface is
unrealistic and unnatural in the context of relativity
\cite{Komar}. This is because no information can be obtained from
space-time points which are separated by space-like distances. In
particular, an observer has access only to information originated
from his past light cone \cite{Dautcourt}. This is an immediate
consequence of the laws of relativity, if we assume that physical
observations are made by a single localized observer
\cite{hughhossein}. Second, there has been considerable success in
using null boundaries to formulate the canonical theory of
gravitational radiation on outgoing null surfaces. This is because
in electromagnetism and gravitation (which are mediated by
particles with zero mass), fields propagate in null directions and
along null hypersurfaces \cite{Penrose}. Third, in some
cosmological models of interest, space-time is not globally
hyperbolic and so there are no Cauchy hypersurfaces on which to
specify boundary data. In such cases, data specified on a
space-like hypersurface cannot be used to generate a unique
classical solution and therefore cannot be used to label a
particular point in the phase space. Even if the space-time is
globally hyperbolic, it may not be possible for localized
observers to gather all the necessary boundary data from a
space-like Cauchy hypersurface. Indeed, unless the space-time is
deterministic, there will be no event whose casual past contains
the hypersurface. In this case, no localized observer will have
access to enough data to distinguish between different classical
solutions - i.e. between different elements of the phase space.
Forth, the formulation of gravitational radiation field on the
null surface lays bare the dynamical degrees of freedom in the
theory and allows one to analyze the properties of the
gravitational radiation field in terms of these quantities
\cite{Sachs}\cite{Ellis}\cite{Bondi}.

 In addition, the approach of setting the final data on a null hypersurface is
 essential if we are interested in a theory such as quantum theory
 that observations made by a single localized observer who can collect
 observational data only from that subset of space-time which lies in the
 causal past\cite{hughhossein}.

In this paper in section two, the hamiltonian formulation of
unimodular gravity is developed.As a product a time variable has
been emerged from the theory that can be regarded as a
cosmological time variable.

In section three, a discussion of Dirac observables in general
relativity is given.In addition, Rovelli's constants of motion
\cite{Rovelli} have been introduced. Section four introduces a set
of observables for the theory on the past light cone of a single
localized observer. These observables are similar to Rovelli's
constants of motion on null hypersurfaces. The evolution of these
observables is with respect to time variable obtained from
unimodular theory of gravity.

\section{The unimodular gravity and emerge of time}

The Einstein-Hilbert action for General Relativity is given by
\be
S[g_{\mu\nu}]=\int |det(g_{\mu\nu})|^{1/2}R[g_{\mu\nu}]d^4x.
\label{eq:Hilbertaction}
 \ee
where $R$ be the Riemann scalar computed from the metric tensor
$g_{\mu\nu}$ .

The equations of motion for unimodular gravity can be obtained by
varying the Hilbert action (\ref{eq:Hilbertaction}) subject to the
unimodular coordinate condition,
\be
-|det(g_{\mu\nu})|^{1/2}+1=0.
 \ee

The theory is then equivalent to General Relativity with an
unspecified cosmological constant, the latter appearing as a
dynamical variable unrelated to any parameters in the action.

 An alternative way to obtain the same theory is to include an extra
term in the Hilbert action, so that the new action is
\be
S[g_{\mu\nu}]=\int e[R-\frac{1}{8}(\bigtriangledown_{\mu}
M^\mu)^2]d^4x,
 \ee
where $e=|det(g_{\mu\nu})|^{1/2}$ . The field equation for
$M^{\mu}$ gives rise to an unspecified cosmological constant in
the action.

In the present paper, our considerations will be based on a
Lagrangian formulation for its relative simplicity. Nevertheless,
we are interest to sketch here how the unimodular assumption
manifests itself in Hamiltonian versions of gravity.

 For this
purpose, we now consider the Hamiltonian formulation of the
theory. We rewrite the action as
 \be
S=\int [eR-\frac{1}{8}e^{-1}(\partial_{\mu} m^\mu)^2]d^4x
 \ee
 where $m^\mu=eM^\mu$  and $\partial_{\mu} m^\mu=\dot{m}+\partial_{i} m^i$ .

 The momentum associated with the dynamical
variable $m^0(x)$ is then
 \be
\pi_0(x)=-\frac{1}{4}(\nabla_{\mu} M^\mu)|_x,
\label{eq:momentumconst}
 \ee
while the momentum associated with the dynamical variables
$m^i(x)$ are
\bea
 \pi_i(x)=0\cdot
\label{eq:momentumicon}
 \eea
Since the action does not explicitly depend on the variables
$\dot{m}^i$, the vanishing momenta  $\pi_i$ are primary
constraints.
\bea
 \pi_i(x)\approx 0\cdot
\label{eq:momentumiconst}
 \eea

 To ensure that these primary constraints are
preserved with time evolution, we also require that
$\dot{\pi}_i(x)\approx 0$, which implies that
  \be
\partial_i \pi_0|_x\approx 0\cdot
\label{eq:secondaryconst}
 \ee
The secondary constraints, eq (\ref{eq:secondaryconst}), ensure
that $\pi_0(x)$ is spatially constant. By substituting $\pi_0(x)$
back into the action one obtains
 \be
S=\int e[R-2\pi_0^2]d^4x\cdot
 \ee
One recognizes that  $\pi_0(x)$ is simply the square root of the
cosmological constant.

The Hamiltonian is obtained from the Lagrangian by a Legendre
transformation and is given by
\be
 H=\int_\Sigma
[N({\cal H}-2\sqrt{{}^{(3)}h}\, \pi_0^2)+N^i{\cal
H}_i+\lambda^i\pi_i-\pi_0(\partial_i m^i)]d^3x,
\label{eq:totalhamiltonian}
 \ee
 where  $\sqrt{{}^{(3)}h}d^3x$   is the measure
associated with the 3-metric $h_{ij}$ on $\Sigma$. The
three-dimensional manifold $\Sigma$ is a submanifold of the
space-time ${\cal M}$. The variables $N,N^i,\lambda^i$ and $m^i$
in Hamiltonian equation are Lagrange multipliers and ${\cal H}$
and ${\cal H}_i$ are the Hamiltonian and momentum given by
 \bea
{\cal H}_i(x;h_{ij},\pi^{ij})&:=&-2\pi_{i{}|j}{}^j(x) \eea and
 \bea
{\cal H}(x;h_{ij},\pi^{ij})&:=&{\cal
G}_{ijkl}(x,h_{ij})\pi^{ij}(x)\pi^{kl}(x)-|h|^{\frac{1}{2}}(x)R(x,h_{ij})
 \eea
in which
\be
{\cal
G}_{ijkl}(x,h_{ij}):=\frac{1}{2}|h|^{1/2}(x)[h_{ik}(x)h_{jl}(x)+h_{jk}(x)h_{il}(x)-h_{ij}(x)h_{kl}(x)]
\ee
We thus have a new Hamiltonian constraint
\be
   {\cal H}_1={\cal H}-2\sqrt{{}^{(3)}h}\, \pi_0^2\approx 0,
  \ee
instead of ${\cal H}\approx 0$.

It can be shown that the new Hamiltonian constraint, the momentum
constraints and the constraints (\ref{eq:momentumconst}) and
(\ref{eq:momentumiconst}) are all first class.

 It has been shown by Henneaux, Teitelboim \cite{Henneaux} and
separately by Unruh \cite{Unruh1989} that the cosmological
constant may be regarded as the momentum conjugate to a dynamical
variable which may be interpreted as the cosmological time
parameter ,
 \be
T(t)=\int_{\Sigma_t}n_\mu M^\mu \sqrt{{}^{(3)} h}\, d^3 x,
\label{eq:cosmologicaltime}
 \ee
with  $n_\mu$ and $\sqrt{{}^{(3)} h}$ are respectively unit normal
to  $\Sigma_t$ and the square root of determinant of the 3-metric
on $\Sigma_t$. An application of Stokes' theorem shows that $T(t)$
is invariant under  $\delta M^\mu
=\epsilon^{\mu\nu\rho\sigma}\nabla_\nu N_{\rho\sigma}$, as it
should be.

 The equation of motion for  $T(t)$ derived from (\ref{eq:cosmologicaltime}),
\bea
\frac{dT}{dt}=\int_{\Sigma_t}N\sqrt{{}^{(3)} h}\, d^3
x=\int_{\Sigma_t}\sqrt{{}^{(4)}
  g}d^3 x,
\eea
 implies that $T(t)$ is just the 4-volume preceding  $\Sigma_t$ plus
some constant of integration. Integration with respect to $t$,
this means that, the change of the time variable equals the
four-volume enclosed between the initial and final hypersurfaces,
which is necessarily positive.This time variable, $T(t)$ may be
regarded as s cosmological time variable, as it continuously
increasing along any future directed time-like curve
\cite{Hossein}. Therefore one my consider $T$ as a monotonically
increasing function along any classical trajectory and so can
indeed be used to parametrise this trajectory.

\section{Dirac observables in General Relativity}

General Relativity, like many other field theories, is invariant
with respect to a group of local symmetry transformations
\cite{Marolf}. The local symmetry group in General Relativity is
the group Diff (${\cal M}$) of diffeomorphisms of the space-time
manifold ${\cal M}$.

In General Relativity, Dirac observables \cite{Dirac} must be
invariant under the group of local symmetry transformations. The
Hamiltonian constraint and momentum constraint in General
Relativity are generators of the symmetry transformations, and so
a function $\Phi$ on the phase space is a Dirac observable, ${\it
iff}$
 \be
\{\Phi,{\cal H}\}=\{\Phi,{\cal H}_i \}= 0,
 \ee
at all points $x \in {\cal M}$. Such observables are necessarily
constants of motion. They are invariant under local Lorentz
rotations ${\it SO(3)}$ and ${\it Diff}\Sigma$   (as well as ${\it
SO (1, 3)}$).

The above criteria for observables in relativity appear to rule
out the existence of local observables if locations are specified
in terms of a particular coordinate system. Indeed, it might
appear that one would be left with only observables of the form
\be
\Phi=\int \phi(x)\sqrt{-g(x)} d^4 x,
 \ee
where  $\phi(x)$ is an invariant scalar as for example $R$, $R^2$
, $R^{\mu\nu} R_{\mu\nu}$ . While such observables clearly have
vanishing Poisson brackets with all the constraints, they can not
be evaluated without full knowledge of the future and past of the
universe. While this may be deducible in principle from physical
measurements made at a specific time, it is well beyond the scope
of any real experimenter.

However, in reality, observations are made locally. We therefore
ought to be able to find a satisfactory way to accommodate local
observables within General Relativity. In particular, we would
like to be able to talk about observables measured at a particular
time, so that we can discuss their evolution. Local observables in
classical or quantum gravity must be invariant under coordinate
transformations. The difficulty in defining local observables in
classical gravity is that diffeomorphism invariance makes it
difficult to identify individual points of the space-time manifold
\cite{Camelia}.

It is fairly easy to construct observables which commute with the
momentum constraints. Such observables can be expressed as
functions of dynamical variables on the spatial hypersurfaces.
However, according to the Dirac prescription, observables must
also commute with Hamiltonian constraint.

In a slightly different formalism, Rovelli addressed the problem
by introducing a Material Reference System (${\it MRS}$)
\cite{Rovelli}. By ${\it MRS}$, Rovelli means an ensemble of
physical bodies, dynamically coupled to General Relativity that
can be used to identify the space-time points.

In Rovelli's approach, all the frames and all the test particles
are assumed to be material objects. However, to implement the
process and simplify the calculation, one has to neglect the
energy-momentum tensor of matter fields in the Einstein equations,
as well as their contributions to the dynamical equations for
matter fields \cite{Prugovecki}. Of course the price that one has
to pay for this neglect is obtaining an indeterministic
interpretation of the Einstein equations. General Relativity is
then approximate because we disregard the energy-momentum of the
${\it MRS}$ as well as incomplete (because we disregard dynamical
equations of the ${\it MRS}$ \cite{Reisenberger}. However, the
indeterminism here is not fundamental and does not imply that
Dirac determinism is violated \cite{Rovelli}. In fact this
approximation can arise in any field theory and has always been
resolved by considering a limiting procedure in which the rest
masses, charges, etc., of test bodies tends to zero
\cite{Prugovecki1}.

Rovelli's observables can be interpreted as the values of a
quantity at the point where the particle is and at the moment in
which the clock displays the value $t$. However $t$ itself is not
an observable, even though its conjugate momentum is constant
along each classical trajectory.

By introducing a cloud of particles filling space, with a clock
attached to every particle, one can easily generalize the model to
a continuum of reference system particles, in order to get a
complete material coordinate system and a complete set of physical
observables. Rovelli's 'evolving constants of motion' are genuine
Dirac's observables. They are constant of motion since they
commute with Hamiltonian and momentum constraints, while evolving
with respect to the clock time $t$.

Rovelli's observables are functions defined on spatial
hypersurfaces. He assumes the space-time has a topology $\Sigma
\times R$ where $\Sigma$ is a compact spatial hypersurface and R
is the real time. In order to have evolution into the future or
past the spatial hypersurface must be a Cauchy hypersurface. This
makes sense if the underlying space-time is assumed to be globally
hyperbolic.

Perhaps more importantly, the observations collected by the
observers will not generally be accessible to any single observer,
and so Rovelli's approach is not useful if we set a theory of
observations by a single observer.

As discussed, one may fix the initial data on null hypersurfaces
rather than spatial hypersurfaces. In General Relativity it is
natural to work with a foliation of space-time by space-like
hypersurfaces, as this reflects the older Newtonian idea of a
3-dimensional universe developing with time. This seems close to
our experiences and is easy to visualize. Nevertheless, null
hypersurfaces and null directions should be considered in here for
the reasons already discussed in the introduction. In particular
The approach of setting the final data on a null hypersurface is
essential if we are interested in a theory such as quantum theory
that observations made by a single localized observer who can
collect observational data only from that subset of space-time
which lies in the causal past.

\section{ Constants of motion}

In ADM formalism, the space-time ${\cal M }$ is assumed to be
foliated by a coordinate time $t$. Now, suppose that we choose the
foliated 3-geometry,  $\Sigma(t)$ to be observer's past light-cone
and also the space-time contains a future-directed time-like
geodesic  ${\Gamma}$ representing the world-line of an observer.
Also suppose that the 4-volume time variable $T(t)$ defined in
(\ref{eq:cosmologicaltime}) instead of coordinate time $t$ has
been used to label the 3-surfaces and also the future-directed
time-like geodesic ${\Gamma}$ . We also suppose that the metric
$g$ satisfies unimodular Einstein's equations which are assumed to
include a contribution from the cosmological constant. It is then
possible to construct a covariantly defined quantity determined by
field values on $\Sigma_T(t)$
\be
\Phi(\Sigma_T)=\int_{\Sigma_T} \phi(x)\sqrt{{}^{(3)}h(x)} d^3 x,
 \ee
where  $\phi(x)$ is any scalar invariant on $\Sigma_T(t)$
expressible in terms of $h_{ij}$ , $R^i_{jkl}$, $K_{ij}$.($i, j,
k, l$ are spatial indices running from 1 to 3) and their covariant
spatial derivatives.These quantities are called world line
${\Gamma}$-observables  \cite{Hossein}.

The so called ${\Gamma}$-observables then have vanishing poisson
brackets with any Hamiltonian $H$, equation
(\ref{eq:totalhamiltonian}), which generates time translations of
$\Sigma_T(t)$ along ${\Gamma}$. The observables $\Phi(\Sigma_T)$
have vanishing Poisson brackets with the momentum constraints
since they are covariantly defined functions of the variables on
the 3-surfaces $\Sigma_T(t)$ . However, they do not have vanishing
Poisson brackets with the Hamiltonian constraints ${\cal H}_1$,
since the prespecified foliation is not invariant under local time
evolution \cite{Kuchar}.

If we define new quantities,$\Phi_T(\Sigma_T)$ ; the value
$\Phi(\Sigma_T)$ at a certain time $T$, then these quantities have
vanishing Poisson brackets with the integrated Hamiltonian
constraints,$\{\Phi_T(\Sigma_T),\int {\cal H}_1 d^3x \}= 0$ , and
can be called 'evolving constants of motion'. These observables
are not the same as Rovelli's constants of motion in a sense that
they are not genuine Dirac's observables. Similarly, the dynamical
time $T(t)$ in the new labeling of 3-surfaces is not a Dirac
observable. The evolution of these observables is expressed in
terms of the dynamical variable $T$, whose conjugate momenta,
$\pi_0$ is a first class constraint.

In summary we have seen that an explicit time variable has been
emerged from unimodular theory of gravity, interpreted as a
cosmological time, and can be used by observers as a clock to
measure the passage of time. A set of 'evolving constant of
motion' has been constructed by using the dynamical time variable
emerged from unimodualr gravity which set the condition on the
${\Gamma}$-observables.

\section{ Acknowledgement}
I would like to thank Dr hugh luckock for his help in the
achievement of this work.

\end{document}